\documentclass[aps,prl,twocolumn,superscriptaddress,amsmath,amssymb]{revtex4}
\usepackage{graphicx}
\usepackage{color}
\begin{document}
\title{Strong fields induce ultrafast rearrangement of H-atoms in H$_2$O}
\author{F. A. Rajgara} 
\affiliation{Tata Institute of Fundamental Research, 1 Homi Bhabha Road, Mumbai 400 005, India}
\author{D. Mathur}
\email{atmol1@tifr.res.in}
\affiliation{Tata Institute of Fundamental Research, 1 Homi Bhabha Road, Mumbai 400 005, India}
\affiliation{UM-DAE Centre for Excellence in Basic Sciences, University of Mumbai - Kalina Campus, Mumbai 400 098, India}
\author{A. K. Dharmadhikari} 
\affiliation{Tata Institute of Fundamental Research, 1 Homi Bhabha Road, Mumbai 400 005, India}
\author{C. P. Safvan}
\affiliation{Inter-University Accelerator Centre, Aruna Asaf Ali Marg, New Delhi 110 067, India}

\begin{abstract}
H-atoms in H$_2$O are rearranged by strong optical fields generated by intense, 10 fs laser pulses to form H$_2^+$, against prevailing wisdom that strong fields inevitably lead to multiple molecular ionization and the subsequent Coulomb explosion into fragments. This atomic rearrangement is shown to occur within a single 10 fs pulse. Comparison with results obtained with $\sim$300-attosecond long strong fields generated using fast Si$^{8+}$ ions helps establish thresholds for field strength and time required for such rearrangements. Quantum-chemical calculations reveal that H$_2^+$ originates in the $^1$A state of H$_2$O$^{2+}$ when the O-H bond elongates to 1.15 a.u. and the H-O-H angle becomes 120$^o$. Bond formation on the ultrafast timescale of molecular vibrations (10 fs for H$_2^+$) has hitherto not been reported.  
\end{abstract}
\maketitle

Ready availability of femtosecond lasers has fuelled widespread contemporary interest in molecular dynamics in intense optical fields (for a recent compilation of cogent reviews, see \cite{PUILS}, and references therein). Intense fields have magnitudes that match intra-molecular Coulombic ones, typically $\sim$50 V $\AA^{-1}$. Exposing molecules to them ruptures one or more bonds, a consequence of field-induced electron ejection, leaving behind two or more ionic cores that experience strong Coulombic repulsion. The inevitable Coulomb explosion sets the time limit for the ensuing fragmentation. 

Chemical reactions necessitate rearrangement of molecular constituents such that different moieties form new bonds. {\it A priori}, this would be considered very unlikely in the strong-field regime as, by prevailing wisdom, Coulomb explosion would occur on much shorter timescales. However, results we report in the following challenge this wisdom: we observe strong-field-induced rearrangement of two H-atoms in H$_2$O$^{2+}$ that occurs faster than Coulomb explosion, within 10 fs. 

Techniques used to generate intense optical pulses are intricately coupled to those for generating short pulses. Several laboratories have recently succeeded in generating pulses short enough that only a handful of optical cycles constitute a single pulse \cite{Baltuska,Baker,Niikura,Haworth,Bocharova}. Might the use of such ``few-cycle" pulses yield new insights in strong-field molecular dynamics? 

Strong-field molecular dynamics is essentially driven by three processes \cite{PUILS,Mathur1,Mathur2}: electron rescattering, spatial alignment and enhanced ionization. Rescattering pertains to the action of the oscillating optical field on the ionized electron wherein, after ejection, it is accelerated back towards the molecular core half a cycle after its ``birth", causing further ionization. Spatial alignment occurs when the linearly polarized optical field acts on the induced dipole moment. Enhanced ionization is mediated by charge exchange effects that increase ionization propensity when the bond length becomes double or triple its equilibrium value. The ultrashort domain allows considerable simplification in the scheme of things as there is sufficient time only for rescattering to effectively contribute to the overall dynamics \cite{Mathur1,Mathur2,Tong,Alnaser}: the other processes are effectively ``switched off".

But what constitutes an ultrafast pulse? Until recently, 100 fs pulses would have been deemed ``ultrafast", but not so in the context of work reported here. In 100 fs, the field amplitude changes little between successive cycles, implying an adiabaticity in molecular response: electrons in the outermost orbital tunnel ionize before the field increases any further. Adiabaticity implies that the irradiated molecule ceases to be a molecule (it becomes an ion) well before the optical field has attained its peak. In the ultrafast domain, however, the electronic response is not adiabatic if the pulse is short enough that the irradiated molecule survives to higher fields before ionizing: the electron is thus exposed to much higher, rapidly increasing fields, gaining more energy before rescattering. Indeed, in experiments on high harmonic generation in Ar, 25 fs pulses yield harmonics that are many orders higher than with 100 fs pulses of the same intensity, indicating that Ar in the ultrashort field survives to higher intensities - a consequence of the non-adiabatic response of the atomic dipole to the ultrafast rise-time \cite{Christov,Rae,Brabec}. 

In our experiments 10 fs pulses were generated by the recently-introduced filamentation technique \cite{Hauri}. We adopted the variant of using two gas-filled tubes \cite{Dharmadhikari} in tandem (Fig. 1). A typical mass spectrum obtained with water vapour is shown in Fig. 2a. Molecular ionization totally dominates in the 10 fs regime, in contrast to the situation with 40-100 fs pulses \cite{Mathur1} wherein fragments like O$^{q+}$ ($q$=2-5) are obtained at 5-10\% yield levels (normalized to H$_2$O$^+$) at similar intensities (2$\times$10$^{15}$ W cm$^{-2}$). This is a signature of non-adiabaticity in the few-cycle domain \cite{Mathur1}.   We wish to focus here on data presented in Fig. 2b which shows that at 1\% yield, clear signature is obtained for formation of H$_2^+$ ions. We made measurements over the intensity range 2-8$\times$10$^{15}$ W cm$^{-2}$ and at different operating pressures to verify that the H$_2^+$ signal does, indeed, emanate from the laser-molecule interaction, and from a unimolecular process. 2$\times$10$^{15}$ W cm$^{-2}$ appears to be the threshold intensity below which no H$_2^+$ signal is obtained.

\begin{figure}
\includegraphics[width=12cm]{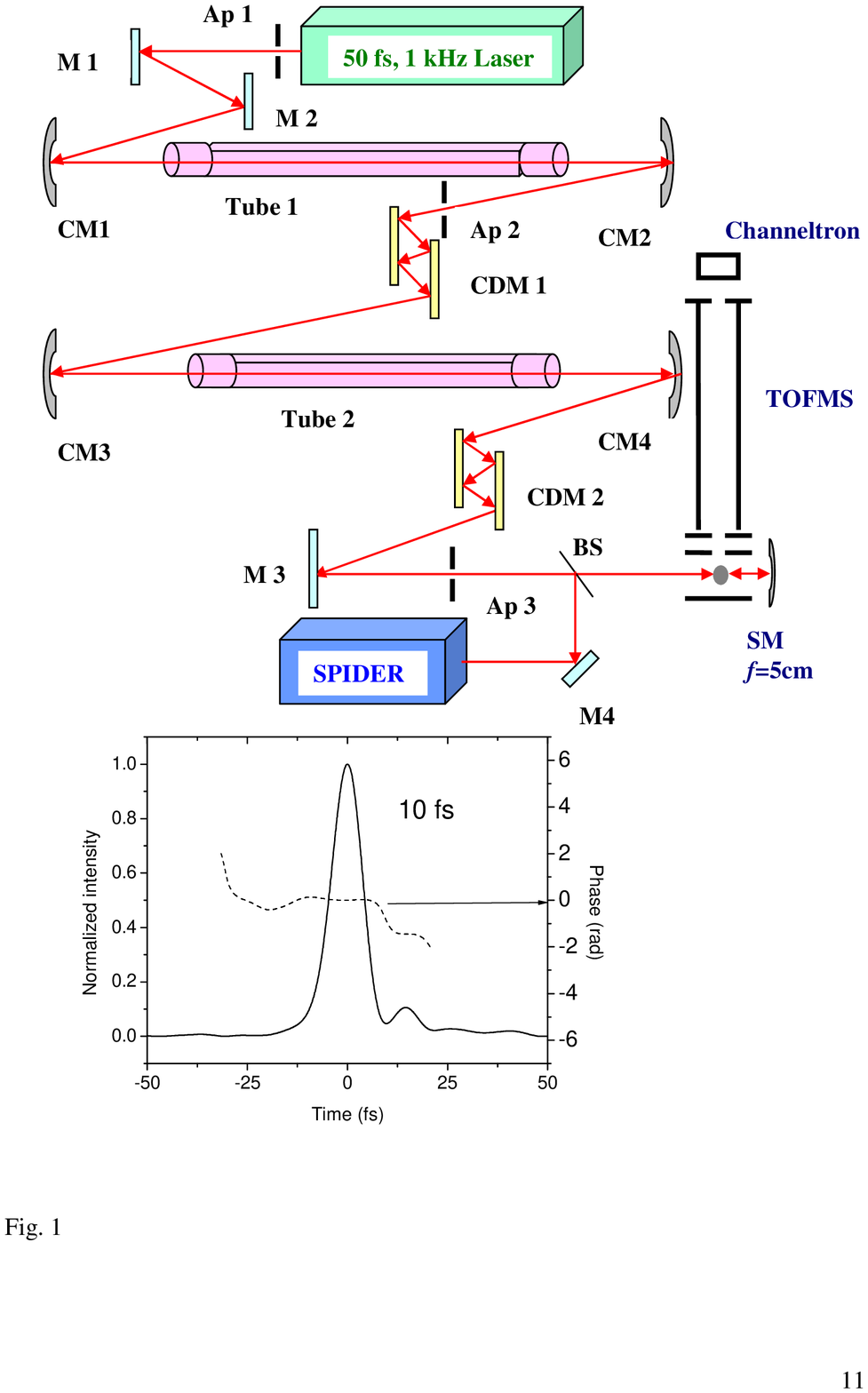}
\caption{Laser light from a Ti-sapphire based amplified system (800 nm wavelength, 0.4 mJ pulses of 50 fs duration at 1 kHz repetition rate) were focused with a metal-coated spherical mirror (CM1, f = 1 m) onto a 1.5 m long tube filled with Ar gas at 1.2 atm where filamentation occurred. The resulting white light output was compressed using a set of chirped dielectric mirrors (CDM1) to produce 16 fs pulses with 0.3 mJ energy which were, in turn, passed through an aperture and focused (using CM4) on to a second 1 m long tube filled with Ar gas at a pressure of 0.9 atm. The resulting broadband light was compressed once again (using CDM2) to yield 10 fs pulses, with typical energy of 0.2 mJ. Temporal and phase features of the pulses were characterized by spectral phase interferometry for direct electric field reconstruction (SPIDER) and steered by sets of high reflectivity mirrors into an ultra high vacuum (UHV) chamber through a 300 $\mu$m thick fused silica window. The laser light was focused within the UHV chamber by a spherical mirror (SM, f=5 cm) to typical peak intensities in the range 10$^{14}$-10$^{15}$ W cm$^{-2}$. Ions formed in the laser-molecule interaction were electrostatically extracted with unit efficiency into a linear (20 cm) time-of-flight spectrometer (TOFMS). A typical SPIDER trace is shown for a 9.7 fs pulse; note the constant, zero phase within the pulse duration.}
\end{figure}

\begin{figure}
\includegraphics[width=12cm]{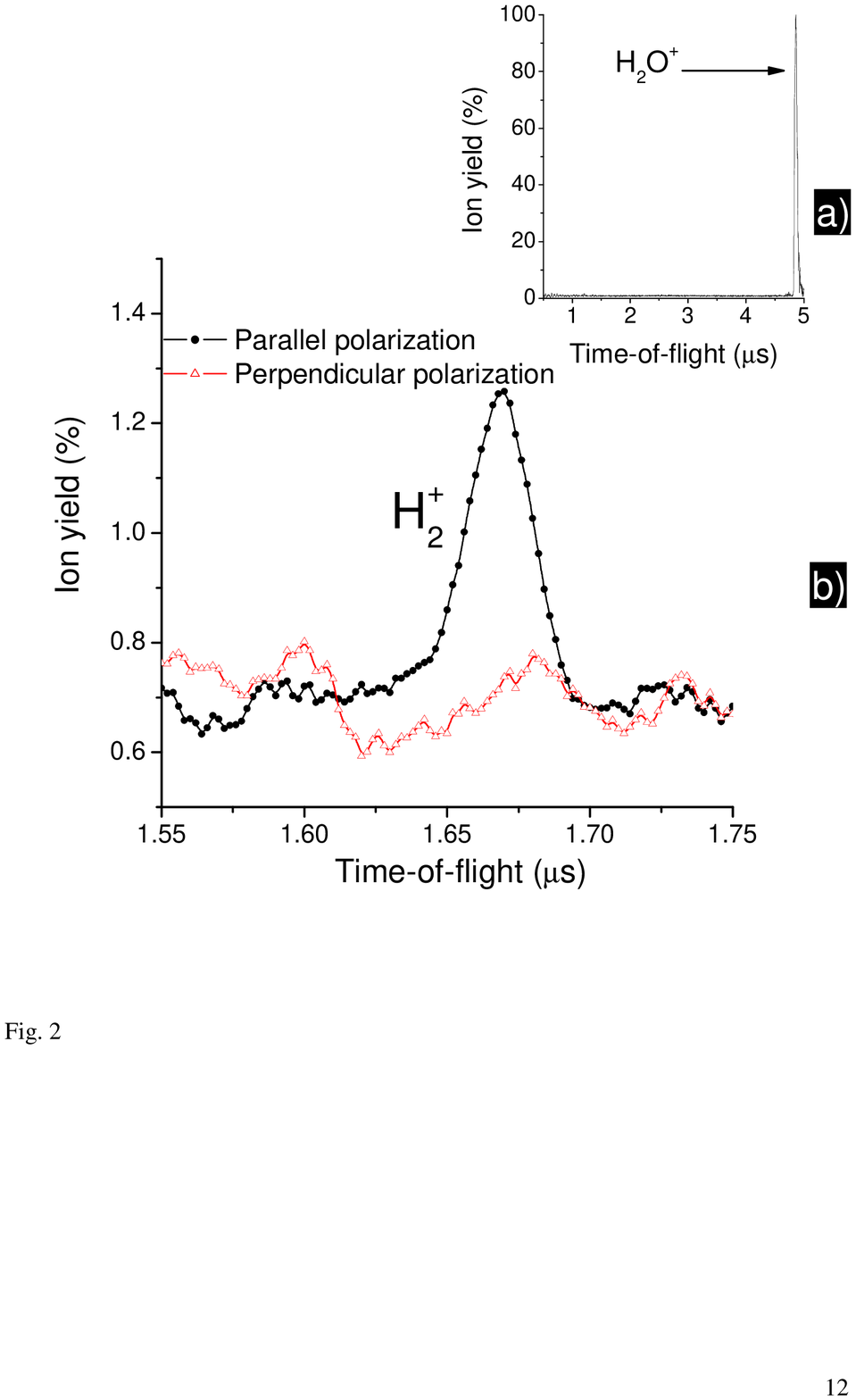}
\caption{a) A typical mass spectrum obtained upon irradiation of water vapour by 10 fs pulses of peak intensity 2$\times$10$^{15}$ W cm$^{-2}$. Note the absence of large-scale fragmentation and the almost total domination of molecular ionization. b) Signal corresponding to H$_2^+$ at yield levels of $\sim$1\% with respect to the H$_2$O$^+$ yield at an intensity of 4$\times$10$^{15}$ W cm$^{-2}$. Note the polarization dependence of this signal: no H$_2^+$ is obtained when the laser polarization vector points in a direction perpendicular to the axis of the time-of-flight spectrometer.}
\end{figure}

The most significant feature of the H$_2^+$ signal is its dependence on the polarization of the ultrashort pulse. H$_2^+$ is obtained only when the polarization vector is parallel to the TOFMS axis; the signal extinguishes when the polarization is orthogonal. This polarization dependence constitutes unambiguous signature of H$_2^+$ being formed by an intramolecular rearrangement. If a bimolecular reaction involving the interaction of a proton with some other water molecule were involved, the process would not exhibit polarization dependence. Moreover, a quadratic dependence on H$_2$O vapor pressure would be expected, contrary to measurements. More significantly, the polarization dependence also indicates that the rearrangement occurs within a single pulse, and is driven by the strong field. If H$_2^+$ formation were simply a consequence of rearrangement in structure of an excited H$_2$O$^+$ state, no polarization dependence would be expected. Taken together, these observations all point to formation of H$_2^+$ ions via motion of two H-atoms on an ultrafast timescale, one that matches the period of O-H vibration. 

Does quantum mechanics allow this? We carried out {\it ab initio} calculations of the potential energy surfaces of neutral and ionized H$_2$O. We made unrestricted Hartree-Fock computations using a 6-311 basis set with two $d$- and one $f$-type of orbitals using the GAMESS suite of programmes \cite{Schmidt}. The $C_n\nu_2$ symmetry of H$_2$O was conserved in all our calculations. An input file was automatically generated at each point of the potential energy surface (with specified O-H distance and H-O-H angle) and calculations were run individually. Part of the potential energy surface of the lowest $^1$A state of H$_2$O$^{2+}$ is shown in Fig. 3a as a contour plot and shows a distinct potential minimum zone when the O-H bond elongates to 1.15 a.u. and the H-O-H angle becomes 120$^o$. Two cuts through this region (shown in Fig 3b for two values of H-O-H angle) indicate that the minimum is deep enough to support a long-lived H$_2^+$ state. We also computed various states of H$_2$O$^{q+}$ ($q$=1-3) but obtained no evidence for a potential minimum capable of supporting H$_2^+$ in any other surface.

\begin{figure}
\includegraphics[width=12cm]{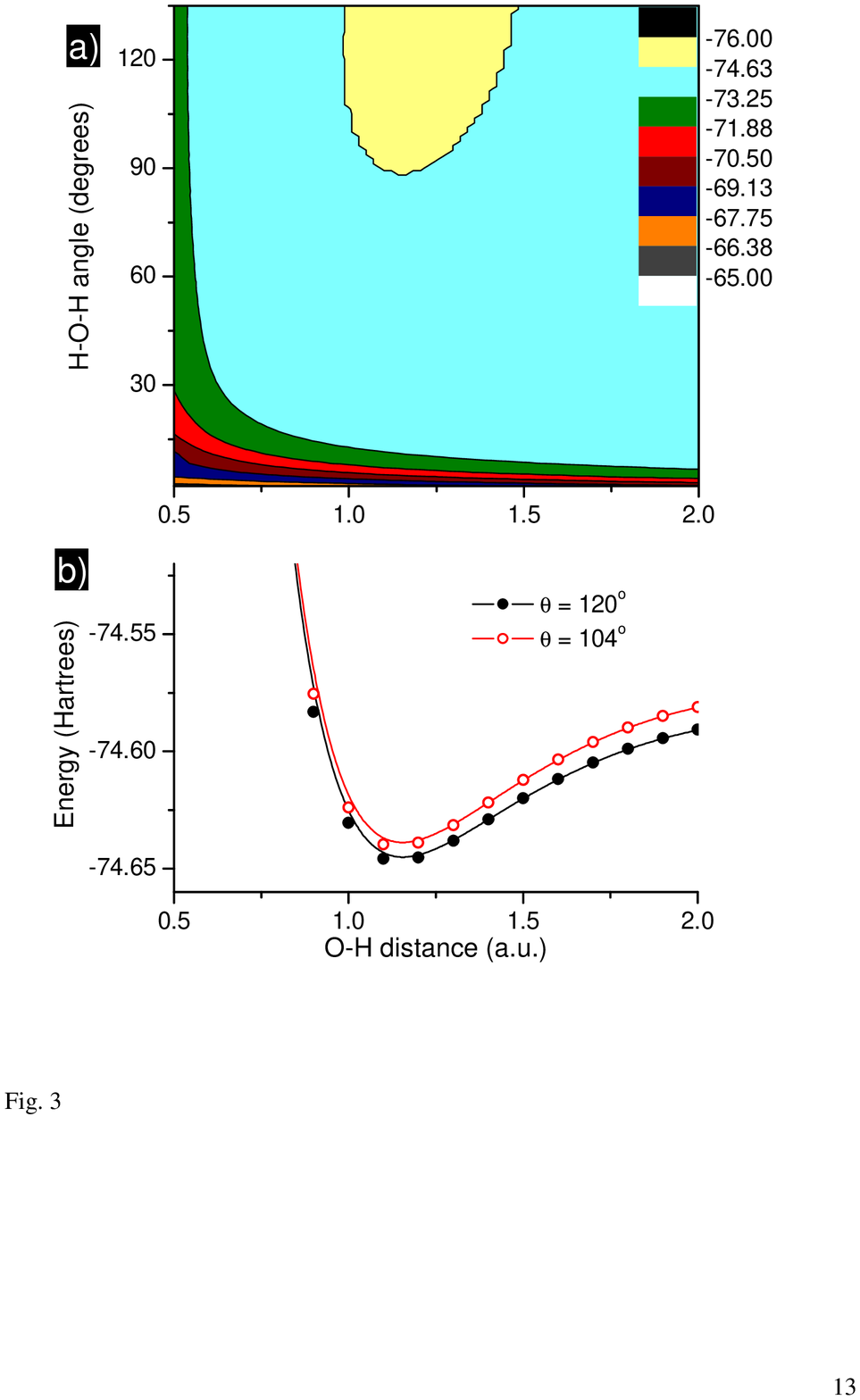}
\caption{a) Contour plot of the potential energy surface of H$_2$O$^{2+}$ in its lowest $^1$A electronic state. Note the zone indicating a potential minimum when the O-H bond length is 1.15 a.u. and the H-O-H angle is 120$^o$. b) Cuts through the potential surface at H-O-H angles of 120$^o$ and 104$^o$ (the equilibrium angle for neutral H$_2$O in the ground electronic state) clearly illustrate the potential minimum.}
\end{figure}

While quantum chemistry appears to theoretically vindicate the formation of H$_2^+$ from doubly-charged water molecules, no time-dependent information is forthcoming from such computations. How fast does a strong-field interaction have to be before such H-H rearrangements become impossible? 

We have attempted to answer this question by conducting experiments wherein a beam of fast, highly-charged ions generates a transient strong field of approximately the same magnitude as our few-cycle laser. Specifically, we made large impact-parameter ($>$3 ${\AA}$) collisions of H$_2$O molecules with 100 MeV Si$^{8+}$ ions from an ion accelerator \cite{Mathur3}. The 8+ charge on the Si-ion ensures a strong field and the energy ensures that the ion-molecule interaction time is only $\sim$300 attoseconds. The resulting ion spectrum (Fig. 4) shows multiply charged fragments, including hydrogen-like O$^{7+}$ at 0.2\% yield levels, but there is no evidence for H$_2^+$ ions. It appears that 300 attoseconds is far too short a time for the two H-atoms to reach the potential minimum indicated in Fig. 3. 

\begin{figure}
\includegraphics[width=12cm]{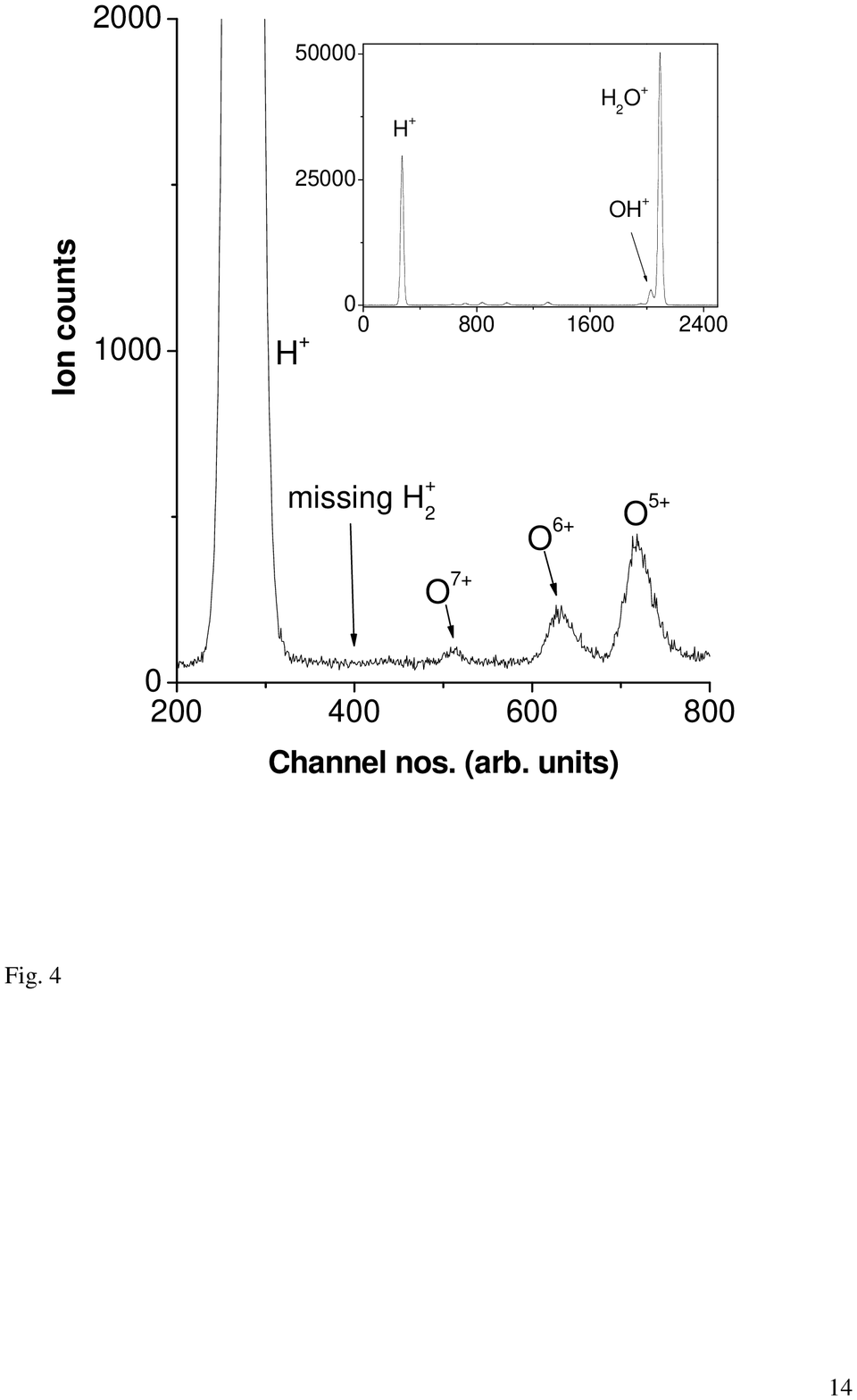}
\caption{Time-of-flight spectrum obtained in collisions of a beam of thermal H$_2$O molecules with 100 MeV Si$^{8+}$ ions from a heavy-ion tandem accelerator. The ion-water interaction time is $\sim$300 attoseconds. Note the absence of a signal corresponding to H$_2^+$ ions. Channel numbers are a measure of flight time (in arbitrary units), with individual ion peaks calibrated with respect to the H$_2$O$^+$ parent ion peak and a peak generated by dopant N$_2^+$ ions. }
\end{figure}

Does this ultrafast phenomenon in water in the strong-field regime have relevance outside the intense laser laboratory? The answer is in the affirmative. Water molecules also experience strong fields in natural, terrestrial environments. Upon freezing water to -150 C ice is formed with hexagonal structure with disordered H-bonds in which each O-atom is located at the centre of a tetrahedron formed by four nearest-neighbour oxygens \cite{Eisenberg}. Following the four-decade-old prescription of Coulson and Eisenberg \cite{Coulson}, water ice with $I_h$ structure can be shown to possess an internal electrostatic field of $\sim$10$^{10}$ V m$^{-1}$. This field corresponds to an effective laser intensity of I$_{eff}\sim$7$\times$10$^{12}$ W cm$^{-2}$. 

By invoking adiabaticity, it becomes mandatory to account for how optical energy transfers to H$_2$O, resulting in formation of H$_2$O$^{q+}$ ($q$>1) which, in turn, leads to the breaking of O-H bonds, leading ultimately to the experimental observable of H$^+$ fragments rearranging to form H$_2^+$. We estimate the energy transfer efficiency using the adiabatic parameter, $\lambda_M$ = $\tau_{laser}$/$\tau_{vib}$, which compares the duration of the laser-molecule interaction, $\tau_{laser}$, with the period of O-H vibration (10 fs) in the water molecule, $\tau_{vib}$. One expects the adiabatic domain when $\lambda_M>>$1. However, for energy transfer from a pulsed field, $\lambda_M$ becomes more meaningful when we account for the duration of the field as well as its peak magnitude with respect to E$_a$, the field magnitude required to suppress the Coulomb barrier to the ionization energy, IE. Taking E$_a\sim$0.25(IE)$^2$ and noting that our laser intensities were in the PW cm$^{-2}$ range, it is clear that our operating conditions were well away from adiabatic. 

In summary, we have probed the strong field ionization of water molecules in the non-adiabatic regime that is accessed when laser pulses of only 10 fs duration are used. We observe a counterintuitive rearrangement of H-atoms on such an ultrafast timescale, manifesting itself in the observation in mass spectra of H2+, a molecular species that is not expected when H$_2$O dissociates.

\acknowledgements
We are grateful to the Department of Science and Technology for partial but important financial support for our femtosecond laser system.

\end{document}